\documentclass[aps,pra,preprint,groupedaddress,showpacs]{revtex4}
\usepackage{amsmath,amssymb,amsthm,amsfonts}
\usepackage{graphicx}
\usepackage{bm}
\begin{document}
\title{Universal description of the rotational-vibrational spectrum of three 
particles with zero-range interactions 
}

\author{O.~I.~Kartavtsev}

\author{A.~V.~Malykh}

\affiliation{ Joint Institute for Nuclear Research, Dubna, 141980, Russia }


\begin{abstract}

A comprehensive universal description of the rotational-vibrational spectrum 
for two identical particles of mass $m$ and the third particle of mass $m_1$ 
in the zero-range limit of the interaction between different particles is 
given for arbitrary values of the mass ratio $m/m_1$ and the total angular 
momentum $L$. 
It is found that the number of vibrational states is determined by the 
functions $L_c(m/m_1)$ and $L_b(m/m_1)$. 
Explicitly, if the two-body scattering length is positive, the number of 
states is finite for $L_c(m/m_1) \le L \le L_b(m/m_1)$, zero for 
$L > L_b(m/m_1)$, and infinite for $L < L_c(m/m_1)$. 
If the two-body scattering length is negative, the number of states is  
zero for $L \ge L_c(m/m_1)$ and infinite for $L < L_c(m/m_1)$. 
For the finite number of vibrational states, all the binding energies are 
described by the universal function 
$\varepsilon_{L N}(m/m_1) = {\cal E}(\xi, \eta)$, where 
$\xi = \displaystyle\frac{N - 1/2}{\sqrt{L(L + 1)}}$,  
$\eta = \displaystyle\sqrt{\frac{m}{m_1 L (L + 1)}}$, and $N$ is 
the vibrational quantum number. 
This scaling dependence is in agreement with the numerical calculations 
for $L > 2$ and only slightly deviates from those for $L = 1, 2$. 
The universal description implies that the critical values $L_c(m/m_1)$ 
and $L_b(m/m_1)$ increase as $0.401 \sqrt{m/m_1}$ and $0.563 \sqrt{m/m_1}$, 
respectively, while the number of vibrational states for $L \ge L_c(m/m_1)$ is 
within the range $N \le N_{max} \approx 1.1 \sqrt{L(L + 1)} + 1/2$.   

\end{abstract}


\maketitle

\section{Introduction}
\label{Introduction}

The universal low-energy few-body dynamics of two-species compounds is of much 
interest both for atomic and many-body physics. 
In this respect, the study of the three-body energy spectrum gives insight 
into the role of triatomic molecules and few-body scattering. 
The area of applications includes the investigation of multi-component 
ultra-cold quantum gases, e.~g., 
binary Fermi-Bose~\cite{Ospelkaus06,Karpiuk05} and 
Fermi~\cite{Shin06,Chevy06,Iskin06} mixtures and of impurities embedded in 
a quantum gas~\cite{Cucchietti06,Kalas06}, which are presently under thorough 
experimental and theoretical study. 
In addition, one should mention the reactions with negative atomic and 
molecular ions~\cite{Penkov99,Jensen03}. 

The universal isotopic dependence of the three-body energy spectrum was 
multiply discussed~\cite{Efimov73,Ovchinnikov79,Li06,DIncao06,Shermatov03}, 
nevertheless, the main objective was the description of Efimov's spectrum. 
Recently, the infinite number of the $1^+$ bound states was 
predicted~\cite{Macek06} for three identical fermions with the resonant 
$p$-wave interaction. 
Concerning the low-energy scattering, one should mention a two-hump structure 
in the isotopic dependence of the three-body recombination rate of 
two-component fermions~\cite{Petrov03,Petrov05a,Kartavtsev07} and 
the two-component model for the three-body recombination near the Feshbach 
resonance~\cite{Kartavtsev02}.

The main aim of the paper is a comprehensive description of the finite 
three-body rotational-vibrational spectrum in the zero-range limit of 
the interaction between different particles. 
Both qualitative and numerical results are obtained by using the solution 
of hyper-radial equations (HREs)~\cite{Macek68,Kartavtsev99,Kartavtsev06}. 
The detailed study of the bound states and scattering problems for the total 
angular momentum $L = 1$ was presented in~\cite{Kartavtsev07}. 

\section{Outline of the approach}
\label{approach}

Particle 1 of mass $m_1$ and two identical particles 2 and 3 of mass $m$ are 
described by using the scaled Jacobi variables ${\mathbf x} =
\sqrt{2\mu}\left({\mathbf r}_2 - {\mathbf r}_1\right),\ 
{\mathbf y} = \sqrt{2\tilde\mu}[{\mathbf r}_3 - 
(m_1 {\mathbf r}_1 + m {\mathbf r}_2)/(m_1 + m)]$ and the corresponding 
hyper-spherical variables $x = \rho\cos\alpha$, $y = \rho\sin\alpha$, 
$\hat{\mathbf x} = {\mathbf x}/x$, and $\hat{\mathbf y} = {\mathbf y}/y$, 
where ${\mathbf r}_i$ is the position vector of the $i$th particle and 
$\mu = m m_1/(m + m_1)$ and $\tilde{\mu} = m (m + m_1)/(m_1 + 2m)$ are 
the reduced masses. 
In the universal low-energy limit, only the s-wave interaction between 
different particles will be taken into account provided the s-wave interaction 
is forbidden between two identical fermions and is strongly suppressed between 
two heavy bosons in the states of $L > 0$. 
The two-body interaction is defined by imposing the boundary condition 
at the zero inter-particle distance, which depends on a single parameter, 
e.~g., the two-body scattering length $a$~\cite{Kartavtsev07}. 
This type of interaction is known in the literature as the zero-range 
potential~\cite{Demkov88}, the Fermi~\cite{Wodkiewicz91} or 
Fermi-Huang~\cite{Idziaszek06} pseudo-potential, and an equivalent 
approach is used in the momentum-space representation~\cite{Braaten03}. 
The units $\hbar = 2\mu = |a| = 1$ are used throughout; thus, the binding 
energy becomes the universal function depending on the mass ratio $m/m_1$ 
and the rotational-vibrational quantum numbers $L$ and $N$. 
In view of the wave-function symmetry under permutation of identical 
particles, a sum of two interactions between different particles is expressed 
by a single boundary condition at the zero distance between particles $1$ and 
$2$, 
\begin{eqnarray}
\label{bch}
\lim_{\alpha\rightarrow \pi/2}\left[ \frac{\partial }{\partial\alpha} - 
\tan\alpha - \rho \frac{a}{|a|} \right]\Psi = 0 \ . 
\end{eqnarray}

The problem under study is conveniently treated by using the expansion of 
the properly symmetrized wave function, 
\begin{equation}
\label{Psi} \Psi = (1 + S\widehat{P})
\frac{ Y_{LM}(\hat{\mathbf y})}{\rho^{5/2}\sin 2\alpha} \sum_{n = 1}^{\infty}
f_n(\rho)\varphi_n^L(\alpha, \rho) \ ,
\end{equation}  
which leads to the hyper-radial equations for the functions 
$f_n(\rho)$~\cite{Kartavtsev07}. 
Here $\widehat{P}$ denotes permutation of the identical particles 2 and 3, 
$S = 1$ and $S = -1$ if these particles are bosons and fermions, respectively, 
$Y_{LM}(\hat{\mathbf y})$ is the spherical function. 
The action of $\widehat{P}$ on the angular variables in the limit 
$\alpha \to \pi/2$ is given by $\widehat{P}Y_{LM}(\hat{\mathbf y}) \to 
(-1)^L Y_{LM}(\hat{\mathbf y})$ and $\widehat{P}\alpha \to  \omega$, where 
$\omega = \arcsin (1+ m_1/m)^{-1}$. 
The functions $\varphi_n^L(\alpha, \rho)$ in the expansion~(\ref{Psi}) are 
the solutions of the equation on a hypersphere (at fixed $\rho$), 
\begin{equation}
\label{eqonhyp1}
\left[\frac{\partial^2}{\partial \alpha^2} - \frac{L(L + 1)}{\sin^2\alpha}
 + \gamma^2_n(\rho)\right]\varphi_n^L(\alpha,\rho) = 0 \ , 
\end{equation} 
complemented by the boundary conditions $\varphi_n^L(0, \rho) = 0$ and 
\begin{equation}
\label{bconhyp}
 \lim_{\alpha\rightarrow \pi/2}
\left(\frac{\partial}{\partial\alpha} - \rho \frac{a}{|a|} \right) 
\varphi_n^L(\alpha, \rho) = S(-)^L\frac{2}{\sin 2\omega} 
\varphi_n^L(\omega, \rho) \ , 
\end{equation} 
where a set of discrete eigenvalues $\gamma_n^2(\rho)$ plays the role of 
the effective channel potentials in a system of the hyper-radial 
equations~\cite{Kartavtsev07}. 
The functions satisfying Eq.~(\ref{eqonhyp1}) and the zero boundary condition 
are straightforwardly expressed~\cite{Bateman53} via the Legendre function 
\begin{eqnarray}
\label{varphi}
\varphi_n^L(\alpha, \rho) = \sqrt{\sin\alpha} 
Q_{\gamma_n(\rho) - 1/2}^{L + 1/2} (\cos\alpha ) \equiv 
\phi_{L, \gamma_n(\rho)}(\alpha)\ .
\end{eqnarray}
The functions $ \phi_{L, \gamma}(\alpha)$ are odd functions on both variables 
$\gamma$ and $\alpha$ satisfying the recurrent relations 
$\sin\alpha\ \phi_{L + 1, \gamma}(\alpha) = (\gamma - L - 1)\cos\alpha \ 
\phi_{L, \gamma}(\alpha) - (\gamma + L)\phi_{L, \gamma - 1}(\alpha )$,
which follow from those for the Legendre functions.
It is convenient to write $\phi_{L, \gamma}(\alpha) = 
A_{L, \gamma}(\cot\alpha)\sin\gamma\alpha + 
B_{L, \gamma}(\cot\alpha)\cos\gamma\alpha$, where $A_{L, \gamma}(x)$ and 
$B_{L, \gamma}(x)$ are simple polynomials on $\gamma$ and $x$, which are 
explicitly given for few lowest $L$ by $A_{0, \gamma}(x) = 1$, 
$B_{0,\gamma}(x) = 0$, $A_{1, \gamma}(x) = -x$, $B_{1,\gamma}(x) = \gamma$, 
$A_{2, \gamma}(x) = 1 - \gamma^2 + 3x^2$, $B_{2,\gamma}(x) = -3\gamma x$,
$A_{3, \gamma}(x) = 3x(2\gamma^2 - 3 - 5x^2)$, and $B_{3,\gamma}(x) =
\gamma (15x^2 + 4 - \gamma^2)$.  
Substituting~(\ref{varphi}) into the boundary condition~(\ref{bconhyp}) and 
using the identity  $\phi_{L+1, \gamma}(\pi/2) = \frac{\partial 
\phi_{L, \gamma}(\alpha)}{\partial\alpha} \Big|_{\alpha=\pi/2}$ 
one comes to the transcendental equation for $\gamma_n^2(\rho)$, 
\begin{eqnarray}
\label{transeq}
\rho \frac{a}{|a|}\ \phi_{L, \gamma} (\pi/2) = \phi_{L + 1, \gamma} (\pi/2)
 - \frac{2 S (-)^L}{\sin 2\omega}\phi_{L, \gamma}(\omega) \ .
\end{eqnarray}

The attractive lowest effective potential determined by $\gamma_1^2(\rho)$ 
plays the dominant role for the binding-energy and low-energy-scattering 
calculations, while the effective potentials in the upper channels for 
$n \ge 2$ contain the repulsive term $\gamma_n^2(\rho)/\rho^2$ and are of 
minor importance. 
Thus, a fairly good description will be obtained by using the one-channel 
approximation for the total wave function~(\ref{Psi}) where the first-channel 
radial function satisfies the equation~\cite{Kartavtsev07} 
\begin{equation}
\label{system1}
\left[\frac{d^2}{d \rho^2} - \frac{\gamma_1^2(\rho) - 1/4}{\rho^2} + E \right]
f_1(\rho)  = 0 \ .
\end{equation}
Note that the diagonal coupling term is omitted in Eq.~(\ref{system1}), which 
does not affect the final conclusions and leads to the calculation of a lower 
bound for the exact three-body energy. 
Our calculations~\cite{Kartavtsev07} shows that the one-channel approximation 
provides better than few percent overall accuracy of the binding energy.  
 
The most discussed feature~\cite{Efimov73,Ovchinnikov79,Li06,DIncao06} of 
the three-body system under consideration is the infinite number of the bound 
states for small $L$ and large $m/m_1$ (more precisely, for the finite 
interaction radius $r_0$ the number of states unrestrictedly increases with 
increasing $|a|/r_0$). 
As the effective potential in~(\ref{system1}) is approximately given by 
$(\gamma_1^2(0) - 1/4)/\rho^2$ at small $\rho$, the number of vibrational 
states is finite (infinite) if $\gamma_1^2(0) > 0$ ($\gamma_1^2(0) < 0$). 
According to Eq.~(\ref{transeq}), $\gamma_1^2(0)$ decreases with increasing 
$m/m_1$ and becomes zero at the critical value $(m/m_1)_{cL}$. 
Thus, one can define the step-like function $L_c(m/m_1)$, which increases by 
unity at the points $(m/m_1)_{cL}$, so that the number of vibrational states 
is infinite for $L < L_c(m/m_1)$ and finite for $L \ge L_c(m/m_1)$. 
Solving Eq.~(\ref{transeq}) at $\gamma_1 \to 0$ and $\rho \to 0$, one obtains 
the exact values $(m/m_1)_{cL}$, which approximately equal $13.6069657$, 
$38.6301583$, $75.9944943$, $125.764635$, and $187.958355$ for $L = 1 - 5$. 
Originally, the dependence $L_c(m/m_1)$ was discussed in~\cite{Efimov73}. 

Analyzing the eigenvalue equation~(\ref{transeq}) one concludes that for 
$a > 0$ and $S(-)^L = -1$ the effective potential exceeds the threshold energy 
$E = -1$, $\gamma_1^2(\rho)/\rho^2 > -1$, therefore, the bound states only 
exist if either two identical particles are bosons and $L$ is even or two 
identical particles are fermions and $L$ is odd. 
Furthermore, one obtains the trivial answer if $a < 0$ and $L \ge L_c(m/m_1)$, 
for which $\gamma_1^2(\rho) > 0$ and there are no three-body bound states. 

\section{Numerical results}
\label{bound}

The mass-ratio dependence of the binding energies $\varepsilon_{L N}(m/m_1)$ 
for $L \ge L_c(m/m_1)$ and $a > 0$ is determined numerically by seeking 
the square-integrable solutions to Eq.~(\ref{system1}). 
Mostly, the properties of the energy spectrum are similar to those for 
$L = 1$, which were carefully discussed in~\cite{Kartavtsev07}. 
For given $L$, there is the critical value of $m/m_1$ at which the first 
bound state arise, in other words, there are no three-body bound states for 
$L \ge L_b(m/m_1)$, where the step-like function $L_b(m/m_1)$ undergoes 
unity jumps at those critical values. 
Furthermore, all the bound states arise at some values of $m/m_1$ being 
the narrow resonances just below them. 
For the mass ratio near these values, the binding energies and resonance 
positions depend linearly and the resonance widths depend quadratically on 
the mass-ratio excess. 
Exactly at these values one obtains the threshold bound states, whose wave 
functions are square-integrable with a power fall-off at large distances. 
A set of these values of $m/m_1$ (more precisely, the lower bounds for them) 
is obtained numerically and presented in Table~\ref{tab1}. 
With increasing $m/m_1$, the binding energies monotonically increase reaching 
the finite values (shown in Table~\ref{tab1}) at $(m/m_1)_{cL}$; just below 
$(m/m_1)_{cL}$ they follow the square-root dependence on the difference 
$m/m_1 - (m/m_1)_{cL}$. 
Correspondingly, the number of the vibrational states increases with 
increasing $m/m_1$ taking the finite number $N_{max}$ at $(m/m_1)_{cL}$ and 
jumping to infinity beyond $(m/m_1)_{cL}$; in the present calculations 
$N_{max} = L + 1$ for $L < 9$ and $N_{max} = L + 2$ for $10 \le L \le 12$. 
\begin{table}[htb]
\caption{Upper part: Mass ratios for which the $N$th bound state of the total 
angular momentum $L$ arises. 
Lower part: Binding energies $\varepsilon_{L N}$ for the mass ratio fixed at 
$(m/m_1)_{cL}$. } 
\label{tab1}
\begin{tabular}{lccccc}
$N$ & $L = 1$ & $L = 2$ & $L = 3$ & $L = 4$ & $L = 5$ \\
\hline
 1 & 7.9300 & 22.342 & 42.981 & 69.885 & 103.06 \\
 2 & 12.789 & 31.285 & 55.766 & 86.420 & 123.31 \\
 3 & -      & 37.657 & 67.012 & 101.92 & 142.82 \\
 4 & -      & -      & 74.670 & 115.08 & 160.64 \\
 5 & -      & -      & -      & 123.94 & 175.48 \\
 6 & -      & -      & -      & -      & 185.51 \\
\hline
 1 & 5.906  & 12.68  & 22.59  & 35.59  & 52.16  \\
 2 & 1.147  & 1.850  & 2.942  & 4.392  & 6.216  \\
 3 & -      & 1.076  & 1.417  & 1.920  & 2.566  \\
 4 & -      & -      & 1.057  & 1.273  & 1.584  \\
 5 & -      & -      & -      & 1.049  & 1.206  \\
 6 & -      & -      & -      & -      & 1.045  \\
\hline
\end{tabular}
\end{table} 

\section{Universal description of the spectrum}
\label{largeL}

A comprehensive description of the spectrum is obtained by using 
the large-$L$ (correspondingly, large-$m/m_1$) asymptotic expression for 
the binding energies $\varepsilon_{L N}(m/m_1)$. 
Taking the quasi-classical solution of Eq.~(\ref{eqonhyp1}) satisfying 
the zero boundary condition, 
\begin{eqnarray}
\label{qcphi} 
\phi_{L, i\kappa}(\alpha) = \exp\left(\kappa \arccos\frac{x \cos\alpha}
{\sqrt{1 + x^2}}\right) \left(\frac{\sqrt{1 + x^2 \sin^2\alpha} - \cos\alpha}
{\sqrt{1 + x^2 \sin^2\alpha} + \cos\alpha}\right)^{L/2 + 1/4} \ , 
\end{eqnarray} 
where $\gamma_1 = i\kappa$ and $x = \kappa/(L + 1/2)$, 
one writes the eigenvalue equation~(\ref{transeq}) in the form,  
\begin{eqnarray}
\label{qceigenv}
\frac{\rho }{L + 1/2} = \sqrt{1 + x^2} - 
\frac{2 \exp\left(\kappa \arcsin\frac{x \cos\omega}{\sqrt{1 + x^2}}\right)} 
{(L + 1/2)\sin 2\omega} \left(\frac{\sqrt{1 + x^2 \sin^2\omega} - \cos\omega}
{\sqrt{1 + x^2 \sin^2\omega} + \cos\omega}\right)^{L/2 + 1/4} \ . 
\end{eqnarray}
In the limit of large $L$ and $m/m_1$ the eigenvalue equation~(\ref{qceigenv}) 
reduces to 
\begin{eqnarray}
\label{adeigenv}
\rho\cos\omega = u - e^{-u} \ , 
\end{eqnarray}
where $u = \cos\omega\sqrt{\kappa^2 + (L + 1/2)^2}$. 
Notice that taking the limit $\kappa \to 0$ and $\rho \to 0$ 
in~(\ref{adeigenv}) one immediately obtains the relation 
$\cos\omega_{cL} = u_0/(L + 1/2)$, where $\sin\omega_{cL} = 
(m/m_1)_{cL}/[1 + (m/m_1)_{cL}]$ and $u_0 \approx 0.567143$ is the root 
of the equation $u = e^{-u}$; as a result, one finds the asymptotic dependence 
$(m/m_1)_{cL} \approx 6.2179(L + 1/2)^2$ and the inverse relation 
$L_c(m/m_1) \approx (u_0 \sqrt{2m/m_1} - 1)/2 \approx 
0.40103\sqrt{m/m_1} - 1/2$. 
Now the asymptotic dependence $\varepsilon_{L N}(m/m_1)$ 
for large $L$ and $m/m_1$ can be obtained by the quasi-classical solution 
of~(\ref{system1}) with $\gamma_1^2(\rho) = (L + 1/2)^2 - 
[u(\rho)/\cos\omega]^2$ and $u(\rho)$ determined by~(\ref{adeigenv}), 
\begin{equation}
\label{qcint}
\displaystyle\int_{u_-}^{u_+} du \frac{1 + e^{-u}}{u -e^{-u}} 
\sqrt{u^2 - \varepsilon_{L N} (u - e^{-u})^2 - L(L + 1)\cos^2\omega} = 
\pi (N - 1/2)\cos\omega \ ,
\end{equation} 
where $u_-$ and $u_+$ are zeros of the integrand. 

\begin{figure}[hbt]
\includegraphics[width = .8\textwidth]{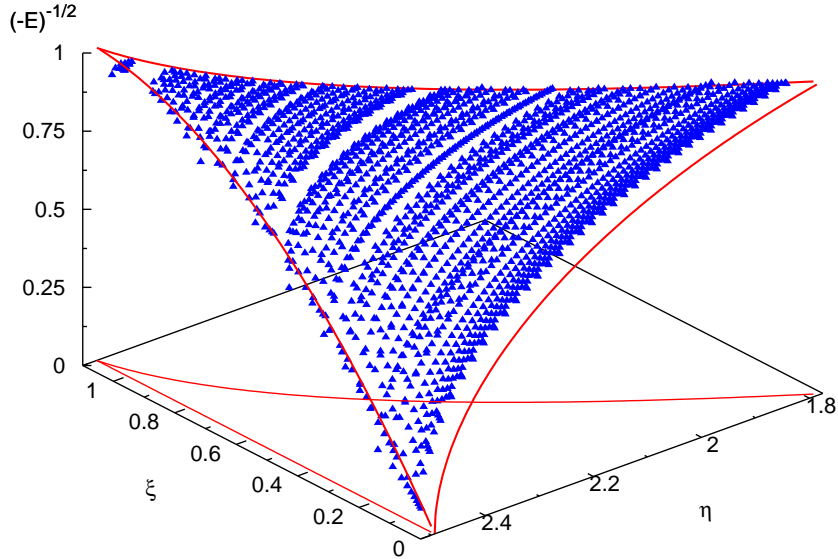}
{\caption{Universal dependence of the bound-state energy $E$ on the scaled 
variables $\xi = \displaystyle\frac{N - 1/2}{\sqrt{L(L + 1)}}$ and 
$\eta = \displaystyle\sqrt{\frac{m}{m_1L(L + 1)}}$. 
The calculated values for $L = \overline{3, 12}$ are plotted by symbols. 
The surface boundary and its projection on the $\xi - \eta$ plane are shown by 
solid lines. 
\label{figen_univ}}}
\end{figure}

Following Eq.~(\ref{qcint}), one expects to express the binding energies via 
the universal function $\varepsilon_{L N}(m/m_1) = {\cal E}(\xi, \eta)$ of two 
scaled variables $\xi = \displaystyle\frac{N - 1/2}{\sqrt{L(L + 1)}}$ and 
$\eta = \displaystyle\sqrt{\frac{m}{m_1 L(L + 1)}}$. 
This two-parameter dependence is confirmed by the numerical calculations 
(up to $m/m_1 \sim 700$), which reveal that the calculated energies for 
$L > 2$ lie on a smooth surface as shown in Fig.~\ref{figen_univ}. 
Even for the smallest $L = 1, 2$ the calculated energies are in good agreement 
with the two-parameter dependence showing only a slight deviation from 
the surface. 

The variables $\xi $ and $\eta $ take values within the area limited by 
the line $\xi = 0$, the line $\eta = \eta_{max} \approx \sqrt{2}/u_0 
\approx 2.493574$ stemming from the condition of finiteness of bound states 
$L \ge L_c(m/m_1)$, and the line ${\cal E}(\xi, \eta ) = 1$ expressing 
the condition of arising of the bound states at the two-body threshold. 
As shown in Fig.~\ref{figen_univ}, the smallest value $\eta = \eta_{min}$ 
is at $\xi = 0$, which corresponds to the condition of arising of the first 
bound state in the large-$L$ limit. 
To find it one requires that $u_+ = u_- \equiv u_b$ at $\varepsilon_{L N} = 1$ 
in Eq.~(\ref{adeigenv}), which leads to $\eta_{min} \approx 
\sqrt{2/(u_b^2 - 1)} \approx 1.775452$, where $u_b \approx 1.278465$ is 
the root of the equation $u = 1 + e^{-u}$. 
This gives the asymptotic dependence for arising of the first bound state, 
$L_b \approx \eta_{min}^{-1}\sqrt{m/m_1} - 1/2 \approx 0.563237 
\sqrt{m/m_1} - 1/2$. 
At the line $\eta = \eta_{max}$ the variable $\xi$ takes its largest value 
$\xi_{max}$, which determines the large-$L$ dependence of the number of 
the vibrational states $N_{max}$ for a given $L$. 
The calculation of the quasi-classical integral~(\ref{qcint}) gives 
$u_- = u_0$, $u_+ \approx 2.872849$, 
$\xi_{max} = \displaystyle\frac{1}{\pi u_0}\int_{u_-}^{u_+} 
\frac{1 + e^{-u}}{u e^u - 1} \sqrt{e^u(2u - u_0^2 e^u) - 1}\ du 
\approx 1.099839 $, and the large-$L$ estimate 
$N_{max} = \xi_{max} \sqrt{L(L + 1)} + 1/2$. 
Taking the entire part of this expression, one can predict that the dependence 
$N_{max} = L + 1$ for $L < 10$ changes to $N_{max} = L + 2$ at $L = 10$, which 
is in agreement with the numerical result. 

The universal surface ${\cal E}(\xi, \eta )$ is bound by three lines, which 
are described by fitting the calculated energies for $L \ge 3$ to simple 
dependencies plotted in Fig.~\ref{figen_univ}. 
As a result, the line defined by ${\cal E}(\xi, \eta ) = 1$ is fairly well 
fitted to $\eta = (\eta_{min} + a\xi)[1 - c \xi (\xi - \xi_{max})]$, where 
$a = (\eta_{max} - \eta_{min})/\xi_{max} \approx 0.652933$ is fixed by 
the evident condition ${\cal E}(\xi_{max}, \eta_{max}) = 1$ and the only 
fitted parameter is $c = 0.1865$. 
Furthermore, the analysis shows that the next boundary line defined by 
$\eta = \eta_{max}$ is described by ${\cal E}^{-1/2}(\xi, \eta_{max}) = 
a_1 \xi(1 - a_2 \xi) [1 - c_1 \xi(\xi - \xi_{max})]$, where 
$a_1 = 1 + u_0 \approx 1.56714$ is fixed by the asymptotic behaviour of 
the integral~(\ref{qcint}) at $L \to \infty$ and $\eta = \eta_{max}$, 
$a_2 \approx 0.38171$ is fixed by the condition 
$a_1 \xi_{max}(1 - a_2 \xi_{max}) = 1$, and the only fitted parameter is 
$c_1 = 0.1881$. 
In particular, at the critical mass ratio the binding energy of the deep 
states in the limit of large $L$ is described by 
${\cal E}(\xi, \eta_{max}) \to (a_1 \xi)^{-2}$, i.~e., 
$\varepsilon_{N L}[(m/m_1)_{cL}] = \frac{L(L + 1)}{(N - 1/2)^2(1 + u_0)^2}$. 
The third boundary line at $\xi \to 0$ is described by the dependence 
${\cal E}^{-1/2}(0, \eta) = a_3 \sqrt{\eta_{max} - \eta} 
[1 + c_2 (\eta - \eta_{min})]$, where $a_3 = 1/\sqrt{\eta_{max} - \eta_{min}} 
\approx 1.18$ is fixed by ${\cal E}(0, \eta_{min}) = 1$ and the only fitted 
parameter is $c_2 = 0.3992$.

\section{Conclusion}
\label{Conclusion}

The presented results complemented by the accurate calculations for 
$L = 1$~\cite{Kartavtsev07} provide in the universal low-energy limit 
a comprehensive description of the rotational-vibrational spectrum of three
two-species particles with the short-range interactions. 
Essentially, all the binding energies are described by means of the universal 
function ${\cal E}(\xi, \eta )$ for those $L_c(m/m_1) \le L \le L_b(m/m_1)$ 
which correspond to the finite number of vibrational states. 
One expects that the universal picture should be observed in the limit 
$|a| \to \infty$, e.~g., if the potential is tuned to produce the loosely 
bound two-body state as discussed in~\cite{Blume05,Kartavtsev06}. 

It is of interest to discuss briefly the effect of the finite, though small 
enough interaction radius $r_0 \ll a$. 
For $L < L_c(m/m_1)$ Efimov's infinite energy spectrum is extremely sensitive 
to the interaction radius $r_0$ and to the interaction in the vicinity of 
the triple-collision point, whereas for $L\ge L_c(m/m_1)$ the binding energies 
depend smoothly on the interaction parameters provided $r_0 \ll a$. 
For this reason, one expects not an abrupt transition from the finite to 
infinite number of bound states for $L = L_c(m/m_1)$ but a smeared off 
dependence for any finite value of $r_0/a$. 

It is worthwhile to mention that arising of the three-body bound states 
with increasing mass ratio is intrinsically connected with the oscillating 
behaviour of the $2 + 1$ elastic-scattering cross section and the three-body 
recombination rate. 
In particular, for $L = 1$ it was shown in~\cite{Kartavtsev07} that two 
interference maxima of the scattering amplitudes are related to the arising of 
two three-body bound states. 
Analogously, the dependence of the scattering amplitudes on the mass ratio for 
higher $L$ would exhibit the number of interference maxima which are related 
to arising of up to $N_{max} = 1.099839 \sqrt{L(L + 1)} + 1/2$ bound states.

Concerning possible observations of the molecules containing two heavy and one 
light particles in the higher rotational states, one should mention 
the ultra-cold mixtures of $^{87}\mathrm{Sr}$ with lithium isotopes 
\cite{Kartavtsev07} and mixtures of cesium with either lithium or helium.
In particular, for $^{133}\mathrm{Cs}$ and $^6\mathrm{Li}$ the mass ratio 
$m/m_1 \approx 22.17$ is just below the value $m/m_1 = 22.34$ at which 
the $L = 2$ bound state arises and $m/m_1 \approx 33.25$ for 
$^{133}\mathrm{Cs}$ and $^4\mathrm{He}$ is above the value $m/m_1 = 31.29$, 
which corresponds to arising of the second $L = 2$ bound state. 
Also, a complicated rotational-vibrational spectrum and significant 
interference effects are expected for the negatively charged atomic and 
molecular ions for which the typical total angular momentum up to $L \sim 100$ 
becomes important due to the large mass ratio. 

\bibliography{fermions}

\end{document}